# Suggested Rules for Designing Secure Communication Systems Utilizing Chaotic Lasers: A Survey

Qingchun Zhao and Hongxi Yin

*Abstract*—Chaotic communications based on semiconductor lasers have aroused great research interest since 1990s. Physical-layer encryption using chaotic lasers is an alternative to transmit message rapidly and confidentially. There are some practical devices and setups for optical chaotic communications, which are intuitively considered to be secure. However, there is lack of a set of security evaluation rules for these communication setups. According to the recent literature, we summarize several criteria for optical chaotic communications to evaluate the security and point out some methods to enhance the security. These criteria and suggested rules are very helpful in designing secure communication systems using chaotic lasers. Finally we propose some possible hot topics on security analysis of optical chaotic communications in future.

*Index Terms*—Chaos, semiconductor laser, optical communication, security, cryptography.

## I. INTRODUCTION

WITH the rapid development of modern information networks, high-speed secure transmission of information becomes very essential. Therefore several secure communication schemes and cryptosystems, such as public key encryption, quantum cryptography and quantum telecommunication, chaotic communications, have been proposed.

Chaos, a random phenomenon, is generated by a deterministic nonlinear system. As chaos is a pseudo-random signal with wide bandwidth and it is unpredictable for a long term, it can be used as carrier to securely hide the confidential message. Compared with the widely used public key encryption, chaotic communications is a hardware-based encryption technique at the physical layer. This communication scheme utilizes the random chaotic waveform generated by deterministic system to transmit message securely. Correct message can be obtained at the receiver side only when the devices and parameters of the receiver are the same as those of transmitter. Otherwise, only random chaotic waveform is received for eavesdropper (Eve) [1].

Semiconductors laser subject to the external perturbation can generate chaotic waveforms. The perturbation can be external optical injection by another laser, optical feedback, or optoelectronic feedback. Semiconductor lasers are advantageous for chaotic communications. It has been demonstrated that the bandwidth of optical chaotic carrier is up to several to tens GHz [2]. The correlation dimension of optical chaotic carrier is greater than 4 [3]. The transmission rate of optical chaotic communication is up to 10 Gbits/s [4]. Moreover, the optical chaotic communications can be compatible with the current fiber-optic communications [5], which can save the cost of lines and equipment.

As to optical chaotic communications, the confidential message with small amplitude is expected to be hidden in the random chaotic carrier in time domain. This kind of message transmission is intuitively considered to be secure. However, there is still no systemic cryptographic analysis of optical chaotic communications for cryptographers to design secure communication systems based on chaotic lasers. The security of optical chaotic communications is comprehensively analyzed in detail in this paper. Some rules are proposed according to the previous literature. Several security measurements of optical chaotic communications are summarized. Some security enhanced methods are outlined. At last, we put forward some hot topics on security analysis of optical chaotic communications in future. This paper is very helpful for designing secure communication systems based on chaotic lasers.

## II. MESSAGE ENCRYPTION

There are three main methods of message encryption using optical chaotic communications [6, 7], whose security will be discussed in the following sections.
- Chaotic masking (CMS): The chaotic carrier is generated by the transmitter laser (TL). The message is directly added with this carrier, as shown in Fig. 1 (a).
- Chaotic shift keying (CSK): The message directly modulates the injection current of the TL. Hence, the TL produces the chaotic carrier with message hidden in it. Please see Fig. 1 (b) below.

This work was supported in part by the National Natural Science Foundation of China (NSFC) under Grant 60772001 and 61071123, Open Fund of State Key Laboratory of Advanced Optical Communication Systems and Networks (Peking University), and Scientific Research Start-up Fund of Dalian University of Technology for Introduced Scholars, China.

The authors are with the Lab of Optical Communications and Photonic Technology, School of Information and Communication Engineering, Dalian University of Technology, Dalian 116023, China (e-mail: hxyin@dlut.edu.cn).

- Chaotic modulation (CMO): The output power of TL is added with the message. Then this mixed signal is sent back to the TL by a feedback loop as a modulation to generate the chaotic carrier. Please see Fig. 1 (c) below.

Note that for all these three methods of message encryption, message extraction is achieved by abstracting the received signal with the chaotic carrier generated by the receiver laser (RL).

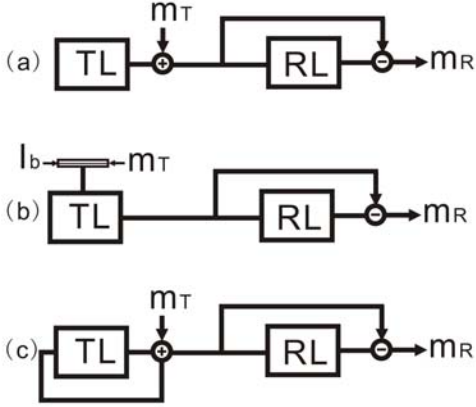

Fig. 1. Message transmission using optical chaotic communications. (a) CMS, (b) CSK, (c) CMO.

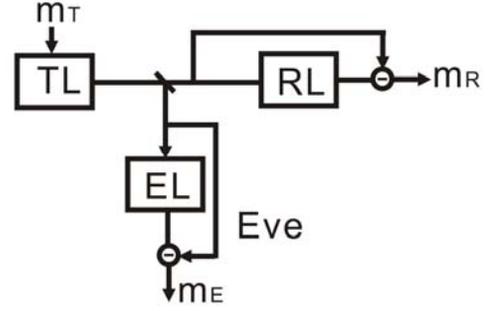

Fig. 2. Breaking optical chaotic communications using generalized synchronization. EL: Eavesdropper laser.

## III. SECURITY MEASUREMENTS

The security issues of optical chaotic communications are mainly related to the followings: different optical chaotic communications; different message encryptions for the identical optical chaotic communication; different message amplitudes, frequencies or phases for the same message encryption and so forth. All of these security differences require some quantitative measurements. This section summarizes the quantitative measurements reported by the previous literature. The corresponding rules for designing secure communication systems based on chaotic lasers are suggested.

### A. Generalized Synchronization

There are two main synchronization schemes for optical chaotic communications, namely complete synchronization and generalized synchronization [8]. For complete synchronization, the parameters between transmitter and receiver are precisely identical, i.e., no parameter mismatch. This synchronization scheme has very good security performance. Eavesdropping is successfully achieved only when Eve has the same laser as the transmitter. Otherwise eavesdropping is unsuccessful. Yet it is impossible to find double lasers with the same parameters, which extremely limits the application of complete synchronization for optical chaotic communications. Alternatively, it is relatively easy to realize generalized synchronization using lasers. This is because a range of parameter mismatch is allowed between the transmitter and receiver under the injection locking. As to Eve, she is also relatively easy to decipher the message. Her laser can synchronize with the transmitter for a range of parameter mismatch (see Fig. 2). Moreover, she can use different types of lasers with transmitter to realize eavesdropping [9]. Hence, the generalized synchronization under the injection locking greatly reduces the security of optical chaotic communications [10]. For generalized synchronization under the injection locking, the security cannot be enhanced even if the feedback length of the external-cavity laser is employed as an additional key [11]. Novel practical synchronization schemes should be paid more attention in future in order to enhance the security of optical chaotic communications.

*Suggested rule 1: The implementation of optical chaotic communications and the security should be taken into account simultaneously.*

### B. Hurst Exponent

Kantelhardt et al. proposed a multifractal degree fluctuation analysis (MFDFA) method to analyze the multifractal characteristics of non-stationary time series, as defined in [12]. Multifractal degree denotes the data complexity. The larger the multifractal degree is, the more complex the structure of the system is. The multifractal degree of the chaotic carrier with message hidden in it is larger than that of the pure chaotic carrier. This method can be used to determine whether the chaotic carrier hides message or not. Zunino et al. demonstrated that Hurst exponent is an effective analyzer of estimating message hidden in chaotic carrier [13]. Their results show that the security of message encryption of CMO is better than CSK. For certain encryption scheme, CMO or CSK, message with large amplitude induces the correlation of the chaotic carrier. Concerning the application, it is difficult for the receiver to accurately extract the message if the optical chaotic carrier hides message with very small amplitude.

*Suggested rule 2: The amplitude of the encrypted message is neither too large nor too small. The appropriate amplitude is about 1% of the average amplitude of the chaotic carrier.*

### C. Nonlinear Filtering

Since chaotic carrier is noise-like, the filtering methods can be applied to examine the security of optical chaotic communications. For these methods the chaotic carrier is considered as the noise. Jacobo et al. studied the security of CMO using the nonlinear filtering method based on the Ginzburg-Landau equation [14]. By adjusting the parameters of the filter, the message can be detected when the averages of the transmitted signals are different for message "0" and "1".

The larger the message amplitude, the better the filtering performance is. Meanwhile this filtering method is ineffective if the averages of the transmitted signals are equal for message "0" and "1".

*Suggested rule 3: The corresponding averages of transmitted signals for message "0" and "1" should be equal to prevent the filtering attack.*

### D. Time-Frequency Analysis

Time-frequency analysis is powerful to detect time-variation and non-stationary signals masked in strong noise. Time-frequency representation provides the joint distribution information in time and frequency domain. Reference [15] defines the mean scalogram ratio and the peak sidelobe level of time-frequency representation to detect the message hidden by chaotic carrier. The efficiency of this method is verified by sinusoidal message. The results shows that the message can be detected if the frequency locates at low power on the power spectrum portrait. Figure 3 shows that the 1-GHz sinusoidal message is very distinct on the time-frequency representation. Compared with CMS, the security performance of CSK is better.

*Suggested rule 4: For message encryption based on optical chaotic communications, time domain, frequency domain, and time-frequency domain should be considered simultaneously in order to hide message securely.*

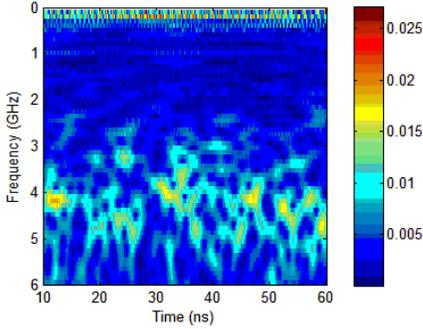

Fig. 3. Time-frequency representation of 1-GHz message encrypted by CSK.

### E. Artificial Neural Network

Artificial neural network establishes the mapping relationship between the output layer and the input one by the transfer functions of the middle layer. Hence, it has a very good ability of learning and memory, which can be utilized for prediction of time series. During the prediction, the time series is divided into training set and prediction one. Firstly the artificial neural network is trained by the training set to assign values for the transfer functions. Secondly the data to be predicted are inputted to the network. At last the output of the network is the forecast value of the time series. According to this principle, the security measurement of optical chaotic communications based on artificial neural network is a known-plaintext attack. For this attack scheme some message and the corresponding transmitted signal in the channel should be obtained by Eve. This attack scheme measures the security of a cryptosystem on condition that part of the message is held by Eve. If Eve cannot break the other part of message, this cryptosystem has a better security performance. Ortín et al. examined the security of optical chaotic communications using this artificial neural network method [16]. Note that the message encryption in their research is CMS.

*Suggested rule 5: To measure the security of optical chaotic communications, the following attack schemes of cryptanalysis can be used: ciphertext-only attack, known-plaintext attack, chosen-plaintext attack, and chosen-ciphertext attack.*

## IV. SECURITY ENHANCEMENTS

### A. Time Delay

There are three schemes for generation of optical chaotic carriers: optical feedback, optoelectronic feedback and optical injection [6]. For all these schemes there exists time delay, for example, the delay induced by the external-cavity. Consequently, the time delay plays an essential role in generating the optical chaotic carrier. Time delay is an extrinsic parameter, which can be used as the ideal key to enhance the security. From the Eve's viewpoint, it is a great success to obtain the time delay just from the output power of TL. On the contrary, the security of optical chaotic communications can be enhanced if the time delay becomes impossible to eavesdrop. In the following context we take the optical feedback scheme as an example to demonstrate cancellation of time delay from the time series of TL.

The relationship between the length of the external-cavity $l$ and the time delay $\tau$ can be expressed by $l = c\cdot\tau/2$ where $c$ is the speed of light in the vacuum. Figure 4 (a) shows the optical chaotic carrier generation using semiconductor laser with a signal external-cavity. These following five methods can be used to extract the signal time delay: return map, autocorrelation function (ACF), average mutual information (AMI), time distribution of extrema, local linear fits in a low-dimensional space [17]. Nevertheless Rontani et al.

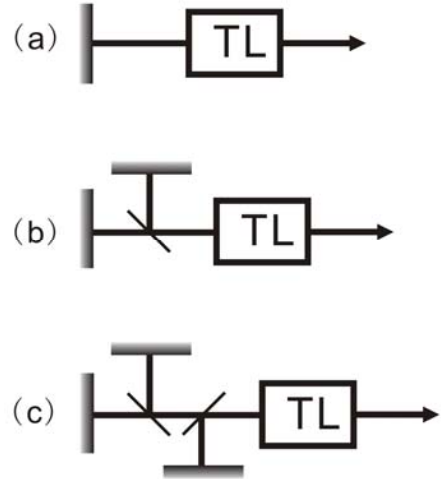

Fig. 4. Time delay of chaotic transmitter. (a) signal delay, (b) double delays, (c) multiple delays.

exhibits their results on the time-delay cancellation from the output power of TL. The time delay is cancelled when the period of relaxation oscillation of TL is close to the delay time by adjusting the feedback rate of the external-cavity and the injection current of TL [18, 19].

*Suggested rule 6: The time delay of TL can be cancelled when the feedback rate of the external-cavity and the injection current of TL are appropriate.*

Semiconductor laser subject to double external cavities exhibits more complex dynamical behaviors because of the additional cavity (Fig. 4 (b)). More adjustable parameters of the external-cavities could be used as keys to enhance the system security [20]. By adjusting the feedback lengths and feedback ratios of the double cavities, Wu et al. experimentally eliminated the time-delay signature of the double-cavity semiconductor laser [21]. Moreover, Tronciu et al. realized the optical chaotic communications exploiting double-cavity semiconductor lasers [22].

*Suggested rule 7: By contrast, double-cavity semiconductor lasers can strengthen the security.*

According to the above discussions, multiple-cavity semiconductor lasers are more secure in optical chaotic communications [23]. In contrast, the structure of the multiple-cavity semiconductor laser is rather complex for practical application. Consequently, there exists a compromise between the system security and the system implementation.

*Suggested rule 8: The multiple-cavity semiconductor lasers are alternative scheme for security enhancement of the chaotic communications.*

Note that the above mentioned external cavities are all fixed. If the external cavities are dynamic, the system security can be strengthened for dynamic key usually replaces the static key in cryptosystem [24]. For Eve, it is more difficult to guess the dynamic key.

*Suggested rule 9: Optical chaotic communications are more secure if the dynamic key is exploited.*

### B. Feedback Phase

External-cavity laser is usually used to generate the carrier for optical chaotic communications. While the short external-cavity laser, sensitive to the change of the feedback phase, has the properties of compactness and integration. Therefore the rapidly changeable feedback phase of the short external-cavity laser can be employed as the key to enhance the security of optical chaotic communications so long as the transmitter and receiver share the same key [25]. For this case, Eve cannot realize eavesdropping even if she equips devices the same as transmitter except the key.

*Suggested rule 10: For the short external-cavity laser, the rapidly changing feedback phase of the external-cavity is an ideal key to increase the security.*

### C. Closed-Loop configuration

There are two configurations for chaotic communications based on the carrier generation by the optical feedback, namely open-loop and closed-loop. When the receiver is a solitary laser with no external-cavity this configuration is open-loop (Fig. 5 (a)). While for the closed-loop configuration the receiver is an external-cavity laser with the symmetric structure as the transmitter (Fig. 5 (b)). Soriano et al. demonstrated that it is easier for the receiver using the closed-loop configuration to extract the message instead of using the open-loop configuration[26]. Hence, the open-loop configuration has worse security. The larger the message amplitude, the easier eavesdropping is, as shown in part B of the above section.

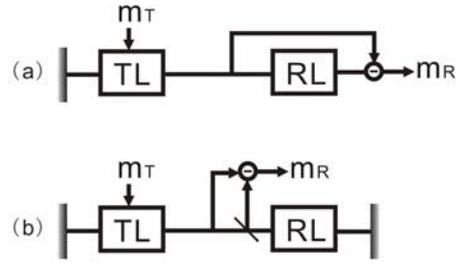

Fig. 5. Schemes for optical chaotic communications. (a) open-loop, (b) closed-loop.

*Suggested rule 11: Compared with the open-loop, the closed-loop configuration for optical chaotic communications is more secure.*

### D. Subcarrier Modulation

The subcarrier modulation technique was applied to the optical chaotic communications to enhance the security by Bogris et al [27]. Figure 6 shows the setup in detail. The message can be successfully encrypted and decrypted when the frequency of the subcarrier locates in the power spectrum of the optical chaotic carrier. Recently the experimental results have been reported [28]. The bit error rate (BER) can reach $10^{-12}$ for transmission of 1 Gbits/s using this setup. Concerning Eve, she cannot achieve eavesdropping even if she knows the structure of TL but have no knowledge of the subcarrier and the frequency $f_{SC}$. Therefore the security is enhanced.

*Suggested rule 12: Subcarrier modulation technique can enhance the security of optical chaotic communications.*

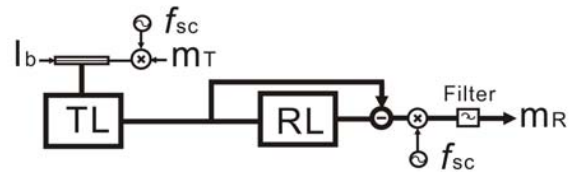

Fig. 6. Schematic of the chaotic transmitter and receiver blocks utilizing subcarrier modulation and demodulation, respectively.

### E. Polarization Mode

Xiang et al. utilized the polarization modes to enhance the security of optical chaotic communications based on the vertical-cavity surface-emitting lasers (VCSELs) [29]. One polarization mode is used as the chaotic carrier to hide the message, while the other orthogonal polarization mode is left as perturbation. A polarizer is arranged at the receiver side to prevent the orthogonal polarization mode into the receiver. The receiver can successfully extract the message for only one polarization mode injects into the receiver. Yet Eve is impossible to break the message assuming that she knows TL but has no knowledge of the polarization modes.

*Suggested rule 13: The polarization mode is a perfect key to enhance the security of optical chaotic communications based on the vertical-cavity surface-emitting lasers.*

## V. CONCLUSIONS

During the past two decades many communication systems based on optical chaos have been proposed in order to transmit message confidentially. However, the security performances of these systems have not attracted enough attention. According to some recent literature, we summarized several measurements to evaluate the security and pointed out some methods to enhance the security. Meanwhile we proposed some rules for designing secure communication systems using optical chaos. The security properties of optical chaotic communications will be a hot topic in future. Here we deem the following issues will attract much research interest:

- more complex topological structure of the transmitter laser to enhance the security
- security of optical communications using spatio-temporal chaos
- security of multiplexing and networks of optical chaotic communications
- security of bidirectional optical chaotic communications.